\documentclass[aps,twocolumn,floats,nofootinbib]{revtex4}
\usepackage{graphics,graphicx,epsfig}
\usepackage{amssymb,color}
\usepackage{epsf,epstopdf}
\usepackage[intlimits]{amsmath}

\newcommand{\beq}{\begin{equation}}
\newcommand{\eeq}{\end{equation}}
\newcommand{\beqn}{\begin{eqnarray}}
\newcommand{\eeqn}{\end{eqnarray}}

\begin{document}

\title{Transcription-dependent spatial organization of a gene locus}

\author{Lev Barinov,$^1$ Sergey Ryabichko,$^{2}$ William Bialek,$^{2,3,4}$ and Thomas Gregor$^{2,3,5}$}

\affiliation{
$^1$Department of Molecular Biology,  $^2$Lewis--Sigler Institute for Integrative Genomics, and $^3$Joseph Henry Laboratories of Physics, Princeton University, Princeton NJ 08544\\
$^4$Initiative for the Theoretical Sciences, The Graduate Center, City University of New York, 365 Fifth Avenue, New York NY 10016\\
$^5$Department of Developmental and Stem Cell Biology UMR3738, Institut Pasteur, 75015 Paris, France
}

\begin{abstract}
There is growing appreciation that gene function is connected to the dynamic structure of the chromosome.  Here we explore the interplay between three--dimensional structure and transcriptional activity at the single cell level. We show that inactive loci are spatially more compact than active ones, and that within active loci the enhancer driving transcription is closest to the promoter.  On the other hand, even this shortest distance is too long to support direct physical contact between the enhancer--promoter pair when the locus is transcriptionally active.  Artificial manipulation of genomic separations between enhancers and the promoter produces  changes in physical distance and transcriptional activity, recapitulating the correlation seen in wild--type embryos, but disruption of topological domain boundaries has no effect. Our results suggest a complex interdependence between transcription and the spatial organization of cis-regulatory elements.
\end{abstract}

\maketitle

It has long been appreciated that developing embryos provide accessible laboratories in which to study the basic mechanisms of gene regulation \cite{sauer+al_96,Zhou+al_97,gregor+al_14}.  More recently it has become clear that embryos provide the opportunity for much more quantitative experiments, illuminating the physics of regulation and the resulting flow of information through the relevant genetic networks \cite{tkacik+al_08, dubuis+al_13, petkova+al_19, dostatniBicoid1, drocco2011, KeenanShvartsman2020}.  Here we exploit the striped expression patterns of the gene {\it even--skipped} ({\em eve}) in the {\em Drosophila} embryo to explore connections between the spatial organization of the chromosome and the functional activation of transcription.

In higher eukaryotes, transcription is controlled by interactions among DNA segments that are widely separated along the chromosome, specifically, cis-regulatory enhancer sequences and their target promoters \cite{deLaat+al_13,furlong+al_18}. These interactions occur within complex genetic loci containing multiple enhancers and in a spatially crowded nuclear environment \cite{misfud+al_15,stadhouders+al_19,vanSteensel+al_19}. It is unclear whether the three-dimensional spatial architecture of such complex loci is connected to their functional output, transcriptional activity \cite{levine+al_14}.  Are there causal links between chromatin conformation and transcription?

Recent studies using imaging and chromatin capture methods  provide conflicting observations regarding locus architecture.
In particular, it is not clear whether physical enhancer--promoter proximity is required for transcription \cite{chen+al_18, cho+al_18, schoenfelder+al_19,benabdallah+al_19, alexander+al_19} or whether   topological domain boundaries are needed to organize the spatial structure of a transcriptionally active locus  \cite{despang+al_2019,williamson+al_2019}.  It seems likely that we are missing something  fundamental about the spatial components of transcriptional regulation. 

\begin{figure*}
\centerline{\includegraphics[trim=0 150 300 200,clip,width = 0.9\linewidth]{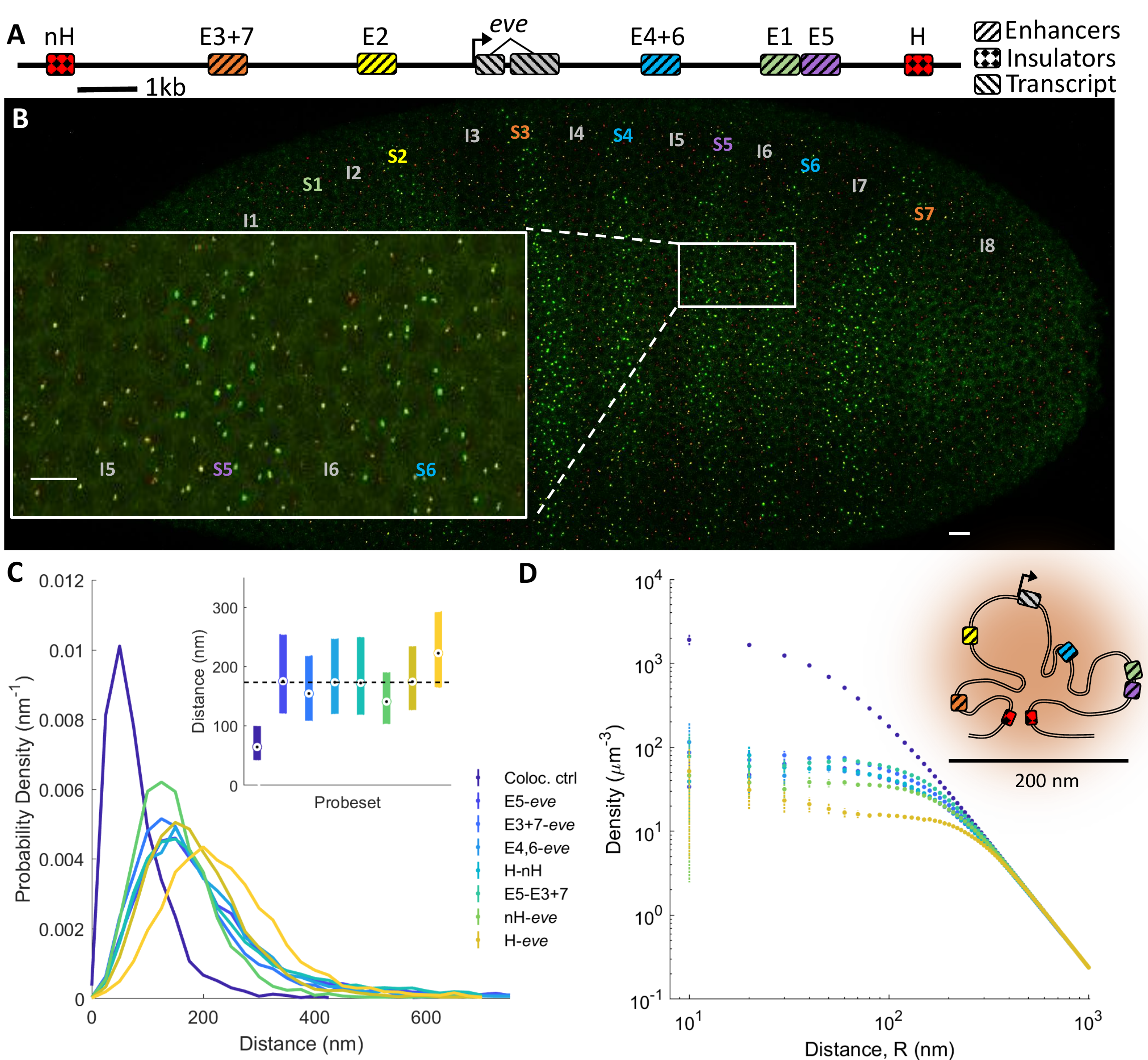}}
\caption{Spatial organization of cis-regulatory elements in the {\it Drosophila} {\it eve} locus.
(A) Schematic sequence of the 18 kb {\em eve} locus with 5 genomically distinct enhancer regions (E1, E2, E3+7, E4+6, and E5), flanked by insulators Homie (H) and nHomie (nH). 
(B) B, Sample image of a 2.5 h old {\it Drosophila} embryo with fluorescent oligonucleotides labeling the {\it eve} promoter in red, the stripe-5 enhancer E5 in yellow, and nascent {\it eve} RNA of the seven-striped pattern (
S1--S7) in green), scale bar $10\,{\rm \mu m}$; inset shows close-up of small white rectangular region, scale bar $10\,{\rm \mu m}$. 
(C) Distributions of distances between pairs of sites, eight combinations of tagged elements: Colocalization control (blue, $n = 6$ embryos,  $N = 4583$ loci), E5 to {\it eve} promoter ($n=23$, $N=6637$), E3+7 to {\it eve} promoter ($n=23$, N=6119), E4+6 to {\it eve} promoter ($n=11$, $N=4222$), Homie to nHomie ($n=21$, $N=5251$), E5 to E3+7 ($n=15$, $N=3612$), nHomie to {\it eve} promoter ($n=17$, $N=5484$), and Homie to {\it eve} promoter (yellow, $n=18$, $N=5053$) from blue to yellow, respectively. Inset shows box and whisker plots for all distributions with 25th, 75th, and median (EC50) values;  average median value is $173\pm 10\,{\rm  nm}$ (dashed line).  (D) Effective density or concentration of sites as a function of the averaging volume (sphere of radius $R$), colors as in (C).  Inset shows  schematic of {\it eve} locus, with all measured points on average at a fixed distance apart, enclosing a droplet-like region (brown shade) with a spatial scale of $\sim\!200\,{\rm nm}$.
}
\end{figure*}

The genomic locus of the {\em Drosophila} patterning gene {\em eve} provides an ideal system in which to address these issues \cite{carroll_90,rivera_96}. {\em Eve} is expressed in the early embryo in seven spatially distinct stripe domains (Fig~1A-B). Each stripe is controlled by one of five enhancers that are all located within an 18 kb topologically associated domain (TAD), which is flanked by two insulator boundaries, {\it homie} and {\it nhomie} \cite{fujioka+al_2009,fujioka+al_2013,hug+al_2017,stadler+al_2017}.  Further, {\em eve} expression occurs in a syncytium prior to the embryo separating into discrete cells. Thus, by observing chromosome conformations in nuclei from different stripe and interstripe domains we have an opportunity to connect spatial structure to transcriptional activity, with all other variables set by the common cytoplasm.

Qualitatively, our strategy is to mark promoter and enhancer sites throughout the {\em eve} locus with fluorescent probes and measure the distances among these sites in large numbers of nuclei across multiple stripes and interstripe domains.  In detail, we use an adapted multi-color oligopaint approach \cite{beliveau+al_2012,cardozo+al_2019}.  We visualize multiple sites within the {\em eve} locus simultaneously using pools of single-stranded DNA (ssDNA) oligonucleotides, each targeting 1--2.5 kb long regions, and Concurrent visualization of DNA and nascent mRNA was performed with a catalyzed reporter deposition scheme. We achieved sub-diffraction resolution with spectrally-based localization microscopy using a conventional confocal microscope to capture optical sections of the surface of blastoderm embryos (Fig~1B). We find that with careful calibration we can localize sites with $\sim\! 25\,{\rm nm}$ precision in the imaging plane and $\sim \! 56\,{\rm nm}$ along the optical axis; our respective optical resolution is $\sim\! 7\,{\rm nm}$, and $\sim\! 55\,{\rm nm}$. As an example, we have tagged two adjacent 1 kb segments with probes in two colors, and find these to be $111\pm 2\,{\rm  nm}$ apart.

To establish the spatial scale of the {\em eve} locus, we measured distances between seven pairs of  elements within the locus, across thousands of nuclei in 15--23 embryos.  We can summarize the results as distributions of pairwise distances (Fig~1C), which have median separations in the range  $140-220\,{\rm  nm}$ and an overall mean of $173\pm 10\,{\rm  nm}$.  The smallest median distance is between the stripe 3+7 (E3+7) and 5 (E5) enhancers, which are separated by 11 kb along the chromosome, while and the largest median distance is between the eve promoter and the  3’ \textit{homie} boundary, 9 kb downstream. These results support a compact globular structure for this 18 kb locus with a spatial scale of $\sim 200\,{\rm  nm}$ (Fig~1D, inset). 

Another way to conceptualize these measurements is to define the probability of a promoter encountering an enhancer element at some distance $R$ from itself. If we normalize this probability by the volume of a corresponding sphere with radius $R$, we can define an effective density or concentration, e.g. for enhancer sites as seen from the promoter (Fig~1D). As expected, if we make $R$ very large this effective concentration decreases, but there is a plateau below $R\sim 200\,{\rm nm}$.  Inside this plateau, it is as if the promoter were in a solution of enhancer sites at a concentration in the range of $20-100\,\mu{\rm m}^{-3}$.  This is equivalent to $30-150\,{\rm nM}$, comparable to the concentration of transcription factors in the surrounding solution \cite{gregor+al_07b,dostatniBicoid1,drocco2011,KeenanShvartsman2020}.  Thus the promoter appears to be no more in contact with enhancers than it would be with the typical {\em unbound} transcription factor.  The spatial proximity of the enhancer and promoter thus does not, by itself, provide a mechanism for transcription factor binding to control transcription.

One might worry that there is a small population of close enhancer--promoter contacts which are invisible to us because of our limited resolution.  As a control we analyze the colocalization of two-color labels targeting a  $\sim$1 kb sequence in Fig~1D (blue), and see that the effective concentration is a factor of 30 to 100 times larger than for any of the enhancer--promoter pairs, plateauing just below $R\sim 20\,{\rm nm}$. This shows that fewer than $1\%$ of the enhancer--promoter pairs could be in close contact, even when the promoter is active, making the possibility of a proximal sub-population unlikely.

Our measurements on fixed tissue are snapshots of a dynamic, fluctuating structure.  We get a more detailed picture of this structure by measuring simultaneously the positions of three sites within the {\em eve} locus, e.g. the enhancers for stripes 3+7 and stripe 5, along with the promoter. In all cases we observe significant cross--correlations among the two pairwise distances, which suggests that the locus forms a fluctuating globule across the whole range of conditions sampled in these experiments.

\begin{figure}
\centerline{\includegraphics[trim=0 200 380 200,clip,width = \linewidth]{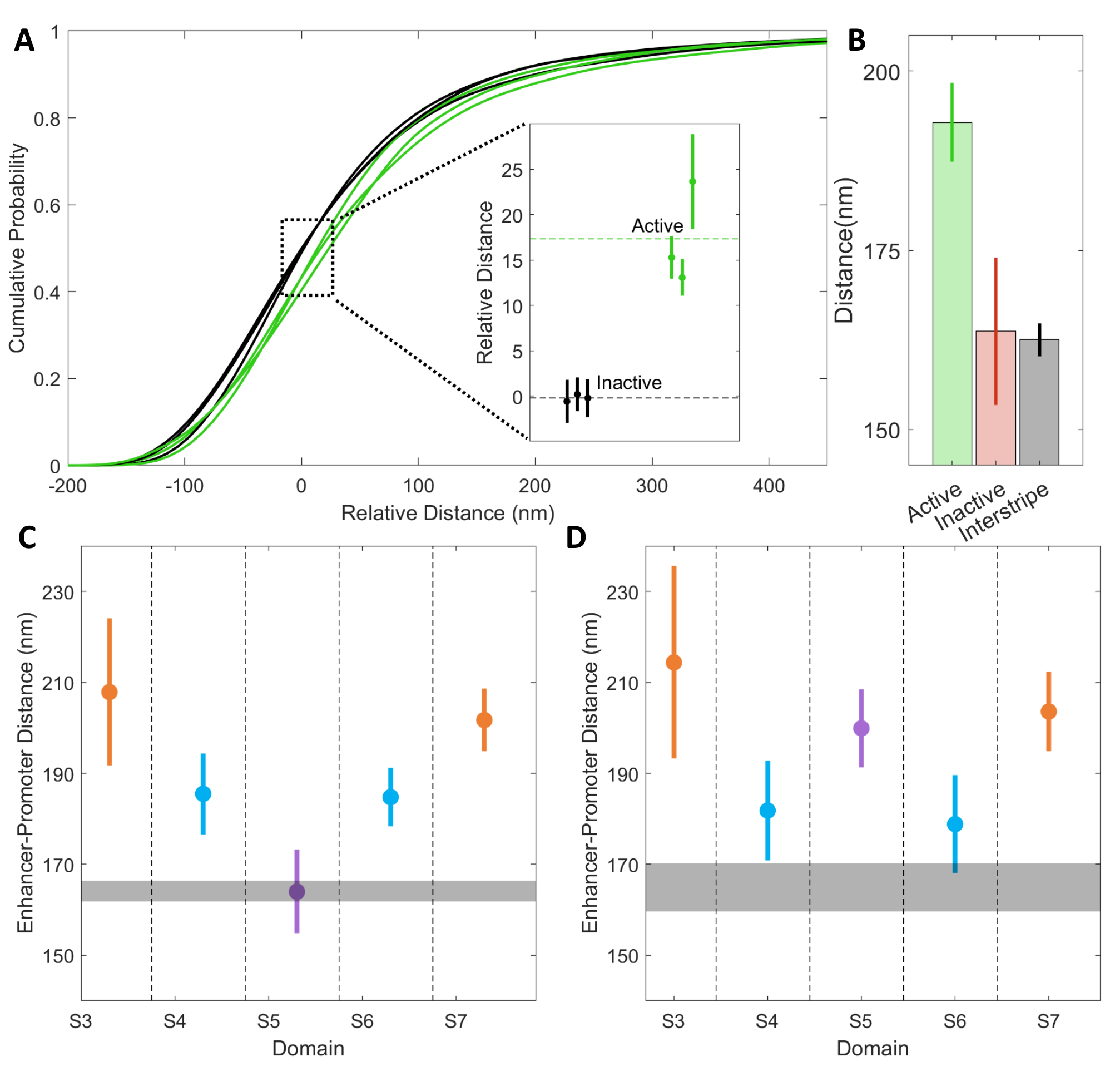}}
\caption{Transcription-dependent and enhancer-specific spatial locus organization. 
(A) Cumulative pairwise enhancer--promoter distance distributions for active (green) and inactive (black) alleles.  Distributions shown for distances measured between the {\it eve} promoter and three enhancers: E5 ($n=23$ embryos, $N_{\rm on}=2471$ loci, $N_{\rm off} =4166$ loci), E4+6 ($n=11$, $N_{\rm on}=697$, $N_{\rm off}=3525$) and E3+7 ($n=23$, $N_{\rm on}=2571$, $N_{\rm off}=3548$). Distances are measured as differences form the median for all inactive loci. Inset shows median distances for inactive (black; average$=0\pm0.4$ nm) and active (green; $17\pm5$ nm) loci.
(B) Pooled median distances of ({\it eve} promoter to E5 and E4+6 ) for active (green, $188\pm9$ nm, $n=34$, $N=204$) and inactive loci; inactive loci are split within-domain (red; $164\pm10$ nm, $n=34$, $N=209$) and inter-domain (gray; $164\pm8$ nm, $n=34$, $N=1699$).
(C) Distances between {\em eve} promoter and E5, in different stripe domains (median$\pm$st. error, $n=23$ embryos). Grey band shows inter-stripe distance ($163\pm4$ nm, $N = 1056$).
(D) As in (C), but for E4+6 enhancer (n=11 embryos; gray band $164\pm11$ nm, N=583).
}
\end{figure}

When we sort nuclei based on whether the {\em eve} locus is transcriptionally active or inactive, we find that the distributions of enhancer--promoter distances are subtly different (Fig~2A).  In particular, median distances  are larger by $\Delta R = 17\pm 5\,{\rm nm}$ at active loci, independent of which enhancer--promoter pair is tagged.  This $\sim\!10\%$ expansion of distances corresponds to a $\sim\!30\%$ expansion of the volume of the locus when it is transcriptionally active, qualitatively in agreement with previous work \cite{benabdallah+al_19}.   We can find inactive loci in the interstripe regions or by searching for alleles within {\em eve}-expressing nuclei that happen to have been captured in a silent state, but the degree of compaction relative to the active state is the same (Fig~2B).  Thus the spatial rearrangements that we see are linked to transcriptional activity, even across alleles in individual nuclei.

\begin{figure*}
\centerline{\includegraphics[width = 0.9\linewidth]{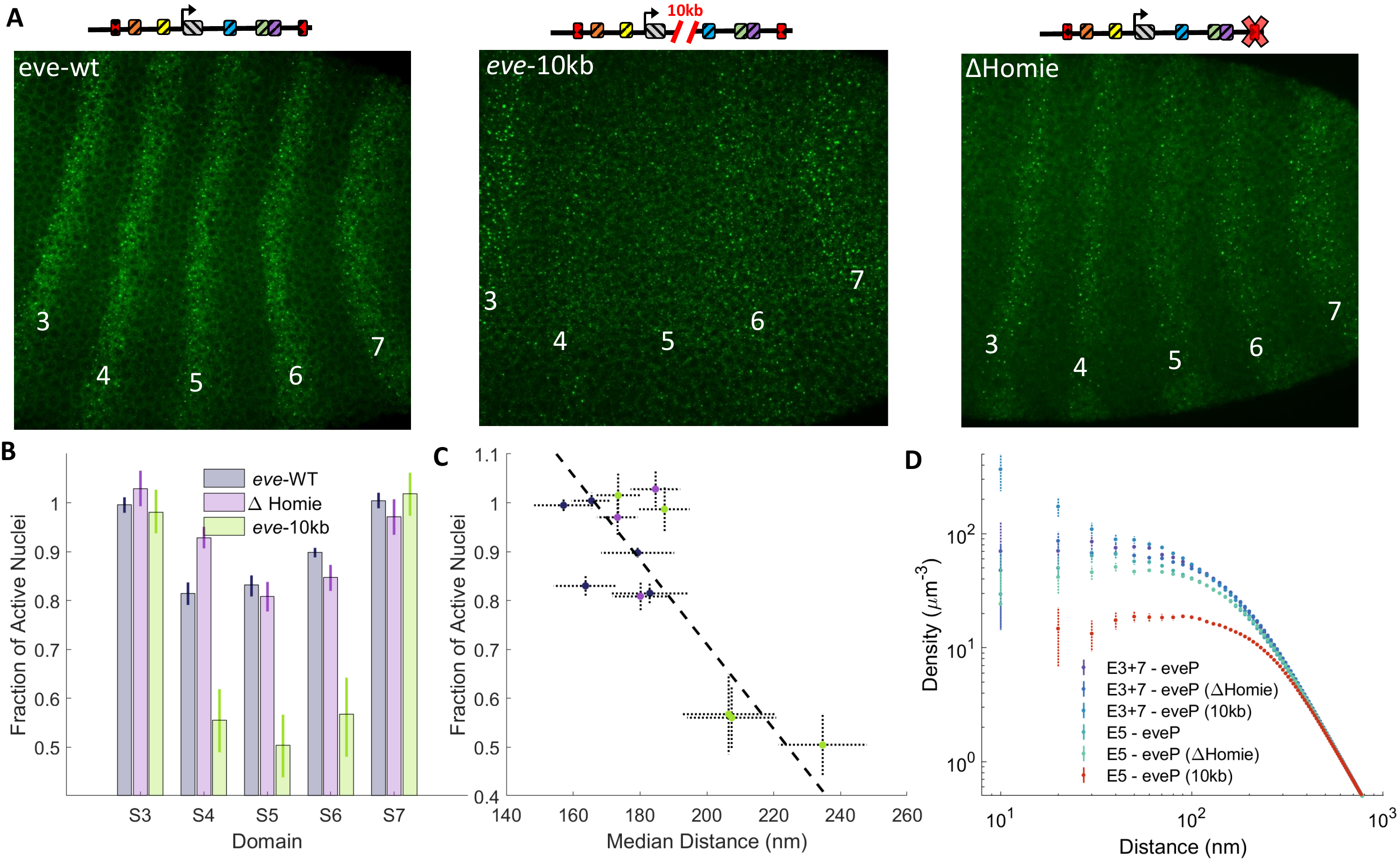}}
\caption{Mutual interaction between spatial locus organization and transcriptional activity. 
(A) Sample images of 2.5 h old embryos with fluorescently labeled {\it eve} transcripts: wild-type (left), an embryo carrying a 10 kb insertion of exogenous DNA (middle), and $\Delta$Homie  (right), with their respective genomic landscapes (top). 
(B) Fraction of active nuclei per stripe domain in WT (blue), $\Delta$Homie (red), and the 10 kb insertion (green), seen in (A) (for each genetic background domain fractions are normalized to average fraction of S3 and S7).
(C) Normalized fraction of active nuclei (from B) versus median distance between the promoter and the relevant domain-driving enhancer the for the three genetic constructs shown in A (same color code as in (B); black dashed line is a linear fit). 
(D) Effective density (for E5 and E3+7) as a function of radial distance (see Fig~1D) for embryos of all three genetic backgrounds shown in A. 
}
\end{figure*}

The fact that different {\em eve} stripes are activated by different enhancers allows us to ask whether enhancer--promoter distances are specific to the interactions that drive transcription. We see that E5 indeed is on average significantly closer to the promoter in stripe 5 (Fig~2C).  Similarly, E4+6 is closer to the promoter in stripes 4 and 6 (Fig~2D).  Interestingly, enhancer--promoter distances are the same in stripes 4 and 6, which are activated by the same enhancer, and similarly in stripes 3 and 7.  These differences in distance are similar to the differences seen between transcriptionally active and inactive loci and the qualitative relationship does not depend on the chosen distance metric. If we think of the individual enhancers as being active or inactive depending on the position of the nucleus, these results can be summarized by saying that enhancers and promoters are close either when both are active or when both are inactive, and farther apart when their states disagree, in qualitative agreement with recent theoretical predictions \cite{bialek+al_19}.  Intriguingly, as an exception to this rule, the E3+7 distance to the promoter appears to be constitutively short, which might be due to additional intra-TAD substructures between the {\it eve} promoter and the upstream insulator element \textit{nhomie}, as evidenced through DNA FISH and Hi-C.
 
Is spatial structure causally related to transcriptional activity? We should be able to change the spatial structure of the {\em eve} locus by inserting inactive segments of DNA, expanding the genomic distances.  We have done this, adding 10 kb of exogenous DNA downstream of the {\em eve} promoter (Fig~3A). This increases the separation between the promoter and the two enhancers E4+6 and E5, while not affecting the separation between the promoter and E3+7 (Fig~1A).  In embryos carrying this modification, we find that the fraction of transcriptionally active nuclei in stripes 4, 5, and 6 is reduced by $40-50\%$ relative to stripes 3 and 7 (Fig~3B). Measurements of enhancer--promoter distances recapitulate the correlations we saw between median distance and activity in wild--type embryos, but now in response to active manipulation (Fig~3C). 

We performed a second manipulation to the locus, deleting a 368 bp DNA segment containing the 3’ {\it homie} insulator element. In embryos carrying this deletion the fraction of active nuclei in the different stripe domains closely matches wild--type levels (Fig~3B). In addition, no major shifts are observed between the WT and $\Delta$Homie spatial configurations when tagging E5 and the promoter (Fig~3C). Thus, the flanking topological boundary has only minor impact on transcriptional activity and spatial organization of the active {\it eve} locus. 

While none of the genomic manipulations impact the E3+7--promoter distance, the additional 10 kb genomic separation between E5 and the promoter leads to a large decrease in transcriptional activity and an increase in physical enhancer--promoter distance, with a corresponding overall decrease in effective enhancer concentration (Fig~3D).
Together these results suggest that a disruption of genomic organization can have a direct impact on both the spatial configuration and the activity level of intervening genes. In particular, there seems to be a strong modulation of transcriptional activity with enhancer--promoter distance at the $150-250\,{\rm nm}$ scale (Fig~2C), encompassing the outer edge of the globular locus structure mapped out in Fig~1D.

In summary, our measurements of enhancer--promoter distances conditioned on transcriptional state solidify a relationship between the spatial organization of the \emph{Drosophila} \emph{eve} locus and transcriptional activity. We find evidence for a globular locus structure with linear dimensions $\sim \!200\,{\rm  nm}$, in line with recent observations of microenvironments  \cite{tsai+al_17}, nuclear hubs \cite{chong+al_18}, or liquid condensates \cite{cho+al_18}. Most strikingly, even transcriptionally active enhancer--promoter pairs are rarely closer than $\sim 150\,{\rm  nm}$, making sustained direct molecular contact highly unlikely, and calling for a novel mechanism for enhancer-to-promoter information transfer \cite{bialek+al_19}.  The plateau we observe in enhancer density for distances below $\sim 200\,{\rm  nm}$ suggests that  enhancer--promoter pairs are insensitive to distance at this length scale, and thus distance cannot be the parameter driving functional control.

Exploiting the multiplicity of the active \emph{eve} domains, each driven by an independent enhancer \cite{small+al_91}, and the high-throughput nature of internally controlled side-by-side experiments, we provide insights into the mechanisms governing both the spatial and transcriptional regulation of the locus at an even finer spatial scale. In agreement with previous measurements \cite{benabdallah+al_19}, we observe a general decompaction of the locus when it is transcriptionally active. In addition, we observe enhancer--specific displacements between enhancer--promoter pairs in the domains in which the respective enhancer is driving activity. These results may help reconcile multiple conflicting observations regarding enhancer--promoter proximity as a function of activity \cite{williamson+al_16,benabdallah+al_19,alexander+al_19,chen+al_18}.  It is especially interesting that enhancers driving transcription are typically the same distance from an active promoter as an inactive enhancer is from an inactive promoter.  

Finally, we demonstrated that the boundaries flanking the {\em eve} TAD are not necessary for enhancer--promoter proximity or for  transcriptional activity. This suggests that  these boundaries may be primarily insulating,  rather than mediating a necessary spatial precursor to transcription. This   is supported by a clear relationship among the enhancer--promoter genomic distances,  their spatial separation, and the transcriptional output. This suggests that the spatial separation of enhancer--promoter pairs is subtly but crucially coupled to  the regulation of transcription, while flanking architectural mediators may not be sufficient to overcome genomic distances on the order of 10 kb within the domains they flank.

We thank MS Levine, M Levo, M Nollmann,  P Schedl, G Tka\v{c}ik, and EF Wieschaus for helpful discussions. This work was supported in part by the US National Science Foundation, through the Center for the Physics of Biological Function (PHY--1734030) and  by National Institutes of Health Grants R01GM097275, U01DA047730, and U01DK127429.


\begin{thebibliography}{99}
%
\bibitem{sauer+al_96}
F Sauer, R Rivera-Pomar, M Hoch, and H J\"ackle, {\it 
Philos Trans R Soc Lond B Biol Sci.} {\bf 351,} 579--87 (1996).
%
\bibitem{Zhou+al_97}
J Zhou, HN Cai, S Ohtsuki, M Levine, {\it Cold Spring Harb Symp Quant Biol.} {\bf 62,} 307--12 (1997).
%
\bibitem{gregor+al_14}
T Gregor, HG Garcia, SC Little, {\it Trends Genet.} {\bf 30,} 364--75 (2014).
%
\bibitem{dostatniBicoid1} 
A Abu-Arish, A Porcher, A Czerwonka, N Dostatni and C Fradin,   {\em Biophys J} {\bf 99,} L33--35 (2010).
%
\bibitem{drocco2011}
JA Drocco, O Grimm, DW Tank and EF Wieschaus, {\em Biophys J} {\bf 101,} 1809--1815 (2011).
%
\bibitem{KeenanShvartsman2020}
SE Keenan, SA Blythe, RA Marmion, NJ-V Djabrayan, EF Wieschaus, and SY Shvartsman,  {\em Dev Cell} {\bf 52,} 794--801 (2020).
%
\bibitem{tkacik+al_08}
G Tkacik, CG Callan Jr, and W Bialek, {\it Proc Natl Acad Sci (USA)} {\bf 105,} 12265--12270 (2008).
%
\bibitem{dubuis+al_13}
JO Dubuis, G Tkacik, EF Wieschaus, T Gregor, and W Bialek, {\it Proc Natl Acad Sci (USA)} {\bf 110,} 16301--16308 (2013).
%
\bibitem{petkova+al_19}
MD Petkova, G Tkacik, W Bialek, EF Wieschaus, and T Gregor, {\it Cell} {\bf176,} 844--855 (2019).
%
\bibitem{deLaat+al_13}
W De Laat and D Duboule,  {\em Nature} {\bf 502,} 499--506 (2013).
%
\bibitem{furlong+al_18}
EEM Furlong and M Levine,  {\em Science} {\bf 361,} 1341--1345 (2018). 
%
\bibitem{stadhouders+al_19}
R Stadhouders, GJ Filion, and T Graf,  {\em Nature} {\bf 559,} 345--354 (2019).
%
\bibitem{vanSteensel+al_19}
B van Steensel and EEM Furlong,  {\em Nat Rev Mol Cell Biol} {\bf 20,} 327--337 (2019).
%
\bibitem{misfud+al_15}
B Mifsud et al,  {\em Nature Genetics} {\bf 47,} 598--606 (2015).
%
\bibitem{levine+al_14}
M Levine, C Cattoglio, and R Tjian,   {\em Cell} {\bf 157,} 13--25 (2014).
%
\bibitem{chen+al_18}
H Chen, M Levo, L Barinov, M Fujioka, JB Jaynes, and T Gregor,   {\em Nat Genet} {\bf 50,} 1296--1303  (2018).
%
\bibitem{schoenfelder+al_19}
S Schoenfelder and P Fraser,  {\em Nat Rev Genet} {\bf 20,} 437--455 (2019).
%
\bibitem{alexander+al_19}
JM Alexander, J Guan, B Li, L Maliskova, M Song, Y Shen, B Huang,  {\em eLife} {\bf 8,} e41769 (2019).
%
\bibitem{benabdallah+al_19}
NS Benabdallah et al. {\em Mol. Cell} {\bf 76,} 1--12 (2019). 
%
\bibitem{cho+al_18}
W-K Cho, J--H Spille, M Hecht, C Lee, C Li, V Grube, and II Ciss\'e,    {\em Science} {\bf 361,} 412--415 (2018).
%
\bibitem{williamson+al_2019}
I Williamson et al {\em Development (Cambridge)} {\bf 143,} 2994-3001 (2016).
%
\bibitem{despang+al_2019}
A Despang, R Sch\"opflin, M Franke,  et al {\em Nat Genet} {\bf 51,} 1263--1271 (2019). 
%
\bibitem{carroll_90}
SB Carroll, {\it Cell} {\bf 60,} 9--16, (1990). 
%
\bibitem{rivera_96}
R Rivera-Pomar and H J\"ackle, {\it Trends Genet.} {\bf 12,} 478--483 (1996).
%
\bibitem{fujioka+al_2009}
M Fujioka, X Wu, and JB Jaynes, {\em Development} {\bf 136,} 3077--3087 (2009).
%
\bibitem{fujioka+al_2013}
M Fujioka, G Sun, and JB Jaynes, {\it PLoS Genetics} {\bf 9} (2013).
%
\bibitem{hug+al_2017}
CB Hug, AG Grimaldi, K Kruse, and JM Vaquerizas, {\it Cell} {\bf 169,} 216--228 (2017).
%
\bibitem{stadler+al_2017}
MR Stadler, JE Haines, MB Eisen, {\it eLife,} {\bf 6} (2017).
%
\bibitem{cardozo+al_2019}
AM Cardozo Gizzi et al {\it Mol. Cell}  {\bf 74,} 212--222 (2019).
%
\bibitem{beliveau+al_2012}
BJ Beliveau, EF Joyce, et al {\em Proc Natl Acad Sci (USA)} {\bf 101,} 21301--6, (2012).
%
\bibitem{gregor+al_07b}
T Gregor, DW Tank, EF Wieschaus, and W Bialek,     {\em Cell} {\bf 130,} 153--164 (2007).
%
\bibitem{bialek+al_19}
W Bialek, T Gregor, and G Tka\v{c}ik, arXiv:1912:08579 [q--bio.sc] (2019).
%
\bibitem{tsai+al_17}
A Tsai, AK Muthusamy, MR Alves, LD Lavis, RH Singer, DL Stern, and J Crocker, {\it Elife} {\bf 6}:e28975 (2017).
%
\bibitem{chong+al_18}
S Chong, C Dugast-Darzacq, Z Liu, P Dong, GM Dailey, C Cattoglio, A Heckert, S Banala, L Lavis, X Darzacq, and R Tjian, {\it Science} {\bf 361,} eaar2555 (2018).
%
\bibitem{small+al_91}
S Small and M Levine, {\it Curr Opin Genet Dev} {\bf 1,} 255--260 (1991).
%
\bibitem{williamson+al_16}
I Williamson, LA Lettic, RE Hill, WA Bickmore, {\it Development} {\bf 143,} 2994--3001 (2016).
%
\end{thebibliography}
\end{document}